\documentclass[12pt]{article}
\usepackage{epsfig}
\newcommand{\bra}[1]{\langle #1|}
\newcommand{\ket}[1]{|#1\rangle}
\newcommand{\braket}[2]{\langle #1|#2\rangle}
\begin{document}
\title{Generalized W-states and Quantum Communication Protocols}
\author{B. Pradhan, Pankaj Agrawal and A. K. Pati\footnote{email:
agrawal@iopb.res.in, bpradhan@iopb.res.in, and akpati@iopb.res.in}\\
Institute of Physics \\ Sachivalaya Marg, Bhubaneswar, Orissa, India 751 005}
\maketitle
\begin{abstract}
We consider W-states and generalized W-states for $n$-qubit systems. We
obtain conditions to use these states as quantum resources to teleport unknown 
states. Only a limited class of multiqubit states can be teleported.
For $one$-qubit states, we use protocols which are simple extensions
of the conventional teleportation
protocol. We also show that these resource states can be used to transmit
{\em at most} $n + 1$ classical bits by sending $n$ qubits if the appropriate conditions are met.
Therefore these states are not suitable for maximal teleportation or superdense
coding when $ n > 3$.
\end{abstract}

\newpage

\section{Introduction}

Entanglement in composite quantum systems allows many quantum information 
processing (QIP) tasks which are otherwise impossible. The power of entanglement
is exploited in a number of quantum communication protocols. Some of the 
examples of these protocols are:  teleportation \cite{bbcjpw}, superdense coding \cite{bw}, 
secret sharing \cite{hbb}, cooperative teleportation \cite{pap}, telecloning \cite{mjpv}, 
quantum cryptography \cite{bb} and many others \cite{hhhh}. 
Both bipartite and multipartite entangled states have been the subject of 
extensive study for their use in these various QIP tasks \cite{hhhh}. Although most of the 
quantum communication protocols involve bipartite systems, multipartite systems
allow many more interesting scenarios. 
Unlike bipartite entanglement which is simple for pure states and quite well 
understood, multipartite
entanglement exhibit rich structure which is far from clear. To explore the facets of
this structure, one can attempt to find out what one may be able to achieve using
multipartite states. In this paper, we are focusing on a class of $n$-qubits pure states.

For three-qubit systems, the states have been classified in two categories 
on the basis of stochastic local operation and classical communication (SLOCC) 
\cite{dvc}. The two categories are GHZ-class and W-class. The states belonging
to these two different categories cannot be converted to one another by SLOCC. 
While the GHZ-state, $\frac{1}{\sqrt{2}}(\ket{000}+\ket{111})$,
 is suitable for the teleportation and the superdense coding,
the W-state, $\frac{1}{\sqrt{3}}(\ket{100}+ \ket{010}) + \ket{001})$,
is not. However, it was shown in \cite{ap}, that a modified
W-state, belonging to the W-class, $\frac{1}{2}(\ket{100}+\ket{010}+\sqrt{2}\ket{001})$,
is suitable for the above tasks. In fact, for
three qubits, the GHZ-state and the modified W-states are specific task-oriented
maximally entangled states (TMES) \cite{pb} for the tasks of teleportation and superdense
coding. However, when one considers more than three-qubit systems and
considers the straightforward generalization of GHZ and W-states, these
states are not TMESs.  For example, GHZ-state could be used for
the tasks of teleportation of one-qubit state and for transmitting $n+1$ 
classical bits by sending $n$-qubits, whereas  a W-state may not always be suitable. We
investigate W-states and generalized W-state as the suitable quantum resources
for these tasks. For our purposes, a generalized GHZ and a 
generalized W-state are similar to the GHZ and the W-state,
except that the coefficients of the superposition can be any complex
parameters. In the next section, 
we consider generalized  W-state and find the conditions on the parameters
of the states for the perfect teleportation and superdense coding. We also
briefly discuss the strategies to teleport {\em genuine} unknown $one$-qubit
state. In the section 3, we consider the 
$n$-qubit W-state and show that an extension of the teleportation 
protocol can help us to teleport a $one$-qubit state, when $n$ is even.
One can also teleport a ${n \over 2}$-qubit state which is a superposition
of two suitable terms.
When $n$ is odd, the W-state is not useful for our tasks.
Last section has some conclusions.

\section{The Generalized W-States}

  We consider  a modified version of the W-state for a $n$-qubit system
 where $n$ can be even or odd. The modified W-state for the $n$-qubits is
\begin{equation}
\ket{WM_{n}}=(a_{1}\ket{1000...0}+a_{2}\ket{0100...0}+a_{3}\ket{0010...0}+...+a_{n} \ket{0000...1}).
\end{equation}
 In this state, the number of terms in the superposition is same as the number of qubits.
 The coefficients $a_{\ell}$ are such that the state is normalized. Therefore,
$  {\cal N}_{n} =  \sqrt{\sum_{\ell = 1}^{n} |a_{\ell}|^2} = 1.$ For convenience,
let us introduce un-normalized $\ket{WM_{n}}$ state, $\ket{WM^{\prime m}_{n}}$. 
It has coefficients $a_m, a_{m+1},....a_n$. For these coefficients
${\cal N}^m_n =  \sqrt{\sum_{\ell = m}^{n} |a_{\ell}|^2} \ne 1.$
The question we wish to address is: What are the possible values of the 
  parameters $a_{\ell}$ for which this state could be suitable for the perfect
  teleportation ? Interestingly this state can always be written as the superposition
  of only two terms
\begin{equation}
\ket{WM_{n}} =  \ket{WM^{\prime 1}_m} \ket{00...0}_{m+1.....n} +            \ket{000...0}_{1.....m}\ket{WM^{\prime m+1}_{n}}.
\end{equation}
Here $m$ can vary between $1$ and $(n-1)$.
In the two terms, the states of $m$-qubit and $(n-m)$-qubit
systems are mutually orthogonal. For such a state, none of the partition
can have entropy more than one. This restricts the classes of the states 
which can be teleported and limits the gains in superdense coding.

One can also define generalized GHZ state --
$(a_1 \ket{000...000} + a_2 \ket{111...111}).$ However as would be clear from
the discussion below, this state would not be suitable for perfect teleportation
or superdense coding, unless coefficients are such that $|a_1|^2 = |a_2|^2 = {1 \over 2}$,
i.e., this state {\em is} GHZ state.

\subsection{Teleportation}

Let us first consider the possibility of teleporting the unknown state of 
$one$-qubit. For this purpose, we write the state $ \ket{WM_{n}}$ as
\begin{equation}
\ket{WM_{n}} =(a_{1}\ket{1000...0}+a_{2}\ket{0100...0}+a_{3}\ket{0010...0}+...+a_{n-1} \ket{0000...1}) \ket{0}_{n} + a_{n}\ket{000...0}\ket{1}_{n},
\end{equation}
or in more compact notations as
\begin{equation}
\ket{WM_{n}} =  \ket{WM^{\prime 1}_{n-1}} \ket{0}_{n} + a_{n}\ket{000...0}\ket{1}_{n}.
\end{equation}
We note that the states $\ket{WM^{\prime 1}_{n-1}}$ and the $(n-1)$-qubit state $\ket{000...0}$
are orthogonal.

   Suppose the unknown state is  $\ket{\psi}_a= \alpha \ket{0}_a+\beta \ket{1}_a$,
 then with the resource state $(3)$, we can write
\begin{eqnarray}
\ket{\psi}_a \ket{WM_{n}} & = & (\alpha \ket{0}_{a} \ket{WM^{\prime 1}_{n-1}} \ket{0}_{n}+ \alpha a_{n} \ket{000...0}_{a12....n-1}\ket{1}_{n}+ \nonumber \\
  & &  \beta \ket{1}_{a} \ket{WM^{\prime 1}_{n-1}} \ket{0}_{n}+ a_{n} \beta \ket{1000...0}_{a12....n-1}\ket{1}_{n} ).                                          
\end{eqnarray}
  Here teleportation would be successful if,
\begin{equation}
           \braket{WM^{\prime 1}_{n-1}}{WM^{\prime 1}_{n-1}} = |a_n|^2.
\end{equation}
   This is because then, we can construct following set of orthogonal measurement
 vectors
\begin{eqnarray}
   \ket{\xi_{1}^{\pm}} & = & \ket{0}_{a} \ket{WM^{\prime 1}_{n-1}} \pm a_{n} \ket{100...0}_{a12....n-1},\nonumber \\
  \ket{\eta_{1}^{\pm}} & = & a_{n} \ket{000...0}_{a12....n-1}  \pm \ket{1}_{a} \ket{WM^{\prime 1}_{n-1}}. 
\end{eqnarray}

    The condition $(6)$ implies that $\sum_{\ell = 1}^{n-1} |a_{\ell}|^2 = |a_n|^2.$
 Using the normalization condition, $ \sum_{\ell = 1}^{n} |a_{\ell}|^2  = 1$,
we obtain
\begin{equation}
         \sum_{\ell = 1}^{n-1} |a_{\ell}|^2 = |a_n|^2 = { 1 \over 2}.
\end{equation}
 This is the condition on the parameters of the generalized W-state.
We note that this condition is equivalent to the statement that the $n$th qubit
is completely mixed and has the entropy one. It is worth pointing out that
any of the qubits of the state $\ket{WM_n}$ can be called $n$th qubit, then
we call the coefficient of the the state $\ket{1}$ of this qubit as $a_n$.
For a three-qubit W-state, the above condition gives the modified W-state
of the Ref \cite{ap}.

  Let us now consider the possibility of teleporting the state of an unknown
two-qubit system. To examine this possibility we rewrite the $\ket{WM_{n}}$
as
\begin{equation}
\ket{WM_{n} } =  \ket{WM^{\prime 1}_{n-2}} \ket{00}_{n-1 n} + \ket{000...0} \ket{\chi^\prime}_{n-1 n},
\end{equation}
where
\begin{equation}
           \ket{\chi^\prime} = (a_{n-1} \ket{10}_{n-1 n} + a_{n}\ket{01}_{n-1 n}).
\end{equation}
   The normalized state would be $\ket{\chi} = \ket{\chi^\prime}/{\cal N}^{n-1}_{n}$,
where ${\cal N}^{n-1}_{n} = \sqrt{|a_{n-1}|^2 + |a_n|^2}$. (Here $\ket{\chi^\prime}
\equiv  \ket{WM^{\prime n-1}_{n}}.$)

   If neither Alice, nor Bob applies multinary transformations to change
the entanglement properties of the state,
 then the state that we could consider for the teleportation can only be
\begin{equation}
        \ket{\psi}_{ab} = \alpha \ket{00} + \beta \ket{\chi}.
\end{equation}
  As in the case of one-qubit state, let us find the conditions on the
 coefficients $a_\ell$ for the possibility of the teleportation of this state. The combined
state is
\begin{eqnarray}
\ket{\psi}_{ab} \ket{WM_{n}} & = & (\alpha \ket{00}_{ab} \ket{WM^{\prime 1}_{n-2}} \ket{00}_{n-1 n}+ \alpha {\cal N}^{n-1}_{n} \ket{000...0}_{ab12....n-2} \ket{\chi}_{n-1 n}+ \nonumber \\
  & &  \beta \ket{\chi}_{ab} \ket{WM^{\prime 1}_{n-2}} \ket{00}_{n-1 n}+  \beta {\cal N}^{n-1}_{n} \ket{\chi}_{ab}  \ket{000...0}_{12....n-2}  \ket{\chi}_{n-1 n}.
\end{eqnarray}
   The condition for this teleportation to succeed is
\begin{equation}
\braket{WM^{\prime 1}_{n-2}}{WM^{\prime 1}_{n-2}} = |{\cal N}^{n-1}_{n}|^2,
\end{equation}
which is equivalent to
\begin{equation}
               \sum_{\ell = 1}^{n-2} |a_{\ell}|^2 = |a_{n-1}|^2 + |a_n|^2.
\end{equation}
This condition together with the normalization conditions give
\begin{equation}
               \sum_{\ell = 1}^{n-2} |a_{\ell}|^2 = |a_{n-1}|^2 + |a_n|^2 = {1 \over 2}.
\end{equation}
In this situation the orthogonal measurement states would be
\begin{eqnarray}
   \ket{\xi_2^{\pm}} & = & \ket{00}_{ab} \ket{WM^{\prime 1}_{n-2}} \pm {\cal N}^{n-1}_{n} \ket{\chi}_{ab} \ket{00...0}_{12....n-2},\nonumber \\
  \ket{\eta_2^{\pm}} & = & {\cal N}^{n-1}_{n} \ket{000...0}_{ab12....n-2}  \pm \ket{\chi}_{ab} \ket{WM^{\prime 1}_{n-2}}. 
\end{eqnarray}
The condition $(15)$ imply that the subsystem of $(n-1)$th and $n$th qubits has
the entropy as one. This again underlines the usefulness of entropy to find
the suitability of a state for teleportation.

We can generalize the above discussion to the teleportation of a multiqubit state.
Because of the structure of the W-state, such a multiqubit state can have only
two terms. When the resource state is a $n$-qubit state then a term
of the teleported state could have at most ${n \over 2}$ qubits (or ${(n-1) \over 2}$
qubits when $n$ is odd). Let us denote this number of qubits by $m$.
 We can rewrite the $\ket{WM_{n}} $ state as
\begin{equation}
\ket{WM_{n}} =  \ket{WM^{\prime 1}_{n-m}} \ket{00.....0}_{n-m+1 n-m+2.... n} + \ket{000...0}_{12....n-m} 
\ket{WM^{\prime n-m+1}_{n}}.
\end{equation}

  One can carry out the steps as above. One will find that the condition for
the teleportation of the state
\begin{equation}
        \ket{\psi}_{a1a2.....am} = \alpha \ket{00....0}_{a1a2....am} + \beta \ket{WM_{m}},
\end{equation}
is
\begin{equation}
\braket{WM^{\prime 1}_{n-m}}{WM^{\prime 1}_{n-m}} = |{\cal N}^{n-m+1}_{n}|^2
\end{equation}

 As above, this implies that the following condition should be met
\begin{equation}
               \sum_{\ell = 1}^{n-m} |a_{\ell}|^2 =  \sum_{\ell = n-m+1}^{n} |a_{\ell}|^2 = {1 \over 2}.
\end{equation}
As earlier, this condition is equivalent to the fact that the subsystem of $m$-qubits
has entropy one.
 The set of orthogonal measurement vectors would be,
\begin{eqnarray}
   \ket{\xi_m^{\pm}} & = & \ket{00....0}_{a1a2....am} \ket{WM^{\prime 1}_{n-m}} \pm {\cal N}^{n-m+1}_{n} \ket{WM_{m}} \ket{00...0}_{1....n-m},\nonumber \\
  \ket{\eta_m^{\pm}} & = & {\cal N}^{n-m+1}_{n} \ket{000...0}_{a1a2....am12....n-m}  \pm \ket{WM^{\prime 1}_{n-m}}_{ab} \ket{WM_{m}}. 
\end{eqnarray}
    We note that when $n$ is even, one suitable resource state can have 
all coefficients $a_{\ell}$ as ${1 \over \sqrt{n}}$. This is the conventional W-state
and we shall discuss it in the next section. Clearly, for odd $n$, there
is no state with ${1 \over \sqrt{n}}$ as coefficients that satisfies 
the above teleportation criteria.

 We now discuss the task of teleporting unknown $one$-qubit state.
If the condition $(8)$ is satisfied, then one of the qubits has entropy one.
If Bob has this qubit and Alice the rest, then as discussed above, the perfect
teleportation is possible. 
     Let us know consider the teleportation of a $one$-qubit state using the 
  resource that is suitable to teleport a state that is a superposition of two
  {\em m}-qubit states. Here $m$ is the number corresponding to the 
   subsystem that satisfies the condition $(20)$, i.e. when
$m$-qubit subsystem has the entropy one. In this case we can write
\begin{eqnarray}
\ket{\psi}_a \ket{WM_{n}} & = & (\alpha \ket{0}_{a} \ket{WM^{\prime 1}_{n-m}} \ket{00...0}_{n-m+1...n}+ \alpha \ket{000...0}_{a12....n-m}\ket{WM^{\prime n-m+1}_{n}}+ \nonumber \\
   &  &  \beta \ket{1}_{a} \ket{WM^{\prime 1}_{n-m}} \ket{00...0}_{n-m+1...n}+  \beta
            \ket{1000...0}_{a12....n-m}\ket{WM^{\prime n-m+1}_{n}}).
\end{eqnarray}
     One can now find a set of orthogonal vectors, as above,
for the measurement. The $one$-qubit state goes over to
a state which has the structure
$ \alpha \ket{00...0}_{12...m} \pm \beta  \ket{WM_{m}},  \beta \ket{00...0}_{12...m} \pm \alpha  \ket{WM_{m}}$. (We have relabeled the subsystem indices as $1,2,...m$.)
 As we shall discuss in more detail
in the next section, one can now adopt at least three strategies.
In the first strategy, the unknown $one$-qubit  state is teleported as the linear
  superposition of two states in a larger Hilbert space. 
One can implement this strategy by constructing
 unitary transformation to apply $\{ \sigma_{0}, \sigma_{1}, i \sigma_{2}, \sigma_{3} \}$
in the subspace $ \{ \ket{00...0},  \ket{WM_{m}} \}$.
In the second strategy, in
  this larger dimensional Hilbert space, one can transfer the unknown qubit state to
  one qubit of Bob by a multinary unitary transformation. 
 To implement this, we can construct unitary operators for
the transformations -- $ \{ \ket{00...0},  \ket{WM_{m}} \} \to \{ \ket{00...0},  \ket{00...1} \}$.
This will reduce the Bob's $m$-qubits to $\ket{00...0}(\alpha \ket{0} + \beta \ket{1})$.
In the third strategy,
  Bob brings a Bell state and makes a multi-particle measurement, followed by
  a unitary transformation to transfer the state to a qubit. This may be called
{\em serial} teleportation. It is possible to implement these strategies.
We shall discuss and 
give the concrete instances of these strategies in the next section. 

\subsection{Superdense Coding}

  In the last subsection, we have seen that  with a suitable choice of parameters, it is
 is possible to teleport a unknown $one$-qubit state using a generalized $n$-qubit W-state. 
 In this situation Alice has $(n-1)$ qubits and Bob has $n$th qubit. If the condition
$(8)$ is satisfied, then this qubit has
 unit entropy. In the case of superdense coding, the distribution
 of the qubits will be reverse. Alice  would have the $n$th qubit and Bob would
 have the remaining $(n-1)$ qubits. 
After  Alice applies the unitary transformations 
$\{\sigma_{0},\sigma_{1},i \sigma_{2},\sigma_{3}\}$ on her qubit, the $\ket{WM_{n}}$
state would become
\begin{eqnarray}
    \sigma_{0} \ket{WM_{n}} & \to & \ket{0}_{n} \ket{WM^{\prime 1}_{n-1}} + a_{n} \ket{1}_{n} \ket{000...0}, \nonumber \\
   \sigma_{1} \ket{WM_{n}} & \to & \ket{1}_{n} \ket{WM^{\prime 1}_{n-1}} + a_{n} \ket{0}_{n} \ket{000...0}, \nonumber \\
  i \sigma_{2} \ket{WM_{n}} & \to & - \ket{1}_{n} \ket{WM^{\prime 1}_{n-1}} + a_{n} \ket{0}_{n} \ket{000...0}, \nonumber \\
  \sigma_{3} \ket{WM_{n}} & \to & \ket{0}_{n} \ket{WM^{\prime 1}_{n-1}} - a_{n} \ket{1}_{n} \ket{000...0}. 
\end{eqnarray}
 We see that if the condition $(8)$ is satisfied, then the four states
  obtained above after unitary transformations would be orthogonal. Therefore
  Bob after receiving the $n$th qubit from the Alice can obtain two cbits.
  We note that in Ref \cite{ap}  the superdense coding protocol was not found to be
  successful in a specific distribution of the qubits. With the proper
  distribution of qubits, Alice with the completely mixed qubit, the superdense
  coding would be possible. Therefore, it is important
  for Alice to use the right qubit to apply the transformation. This qubit has
  to be in a completely mixed state, i.e., with entropy of one. For other distribution of 
  qubits, the superdense coding would not be possible.

 In the more general case, Alice can transmit $(m+1)$ classical bits to
Bob by sending
$m$ qubits. For this to be possible, the state $\ket{WM_n}$ should be such that
the condition $(20)$ is satisfied and Alice should be applying the transformations
on those $m$-qubits that have entropy one. Since maximum possible entropy 
of the subsystem is only one, the maximum gain in the superdense coding is 
also only of one unit \cite{pap} and maximal superdense coding \cite{pb} is
not possible.

\section{ The W-state}

     In this section we shall consider W-state as a special case of
  generalized W-states. However,
     before going over to  the W-state, let us briefly discuss the GHZ-state.
     The $n$-qubit GHZ-state can be written as
\begin{equation}
\ket{GHZ_{n}}=\frac{1}{\sqrt{2}}(\ket{0000...0}+\ket{1111...1}).
\end{equation}
    Each qubit of this state has entropy one. Therefore, Bob can have any qubit and
  Alice can have the remaining qubits. Then by making suitable $n$-qubit measurements
  Alice can teleport an unknown $one$-qubit state. She can also transmit $n +1$ classical
  bits by sending $n$ qubits \cite{pap}. Furthermore, since the subsystems
  of this state with more than one qubit also have unit entropy, one can teleport
  suitable multipartite states with only two terms. 
  However this state is not a TMES with respect to the
  protocols of teleportation and superdense coding for $n > 3$ \cite{pb}.
  The situation is different for a W-state.

The $n$-qubit W-state is given by
\begin{equation}
\ket{W_{n}} = \frac{1}{\sqrt{n}}(\ket{1000...0}+\ket{0100...0}+\ket{0010...0}+...+\ket{0000...1}).
\end{equation} 
 We have found in the last section that this state can be used for teleportation
when $n$ is even. For odd $n$, none of the partitions of this state can satisfy 
the condition $(20)$. When $n$ is even,  Alice and Bob both should have 
${n \over 2}$ qubits each to satisfy this condition. In other words,
for odd number of qubits, all subsystems have entropy less than one
and this state is not suitable for teleportation and superdense coding.
For even number of qubits, only one partition, equal sharing between Alice and
Bob, can have unit entropy. For all other partitions, it is less than one.
Therefore, only this specific partition is useful for teleportation and
superdense coding. 
Because of the symmetry of the state, this would be true for all distribution of qubits.
This is consistent with \cite{lq} where it was found that if subsystems with 
$x$ and $n-x$ numbers of qubits, the entropy of the subsystem is given by 
\begin{equation}  
E_{x|n-x}(W_n)=-\frac{x}{n}\log_2 \frac{x}{n}-(1-\frac{x}{n})\log_2(1-\frac{x}{n}).
\end{equation} 
The entropy attains maximum value of one when $n$ is even and $x$ is $n/2$. For other partitions
the entropy is less than one. 
We shall now examine the W-state with even number of qubits for the purposes of
teleportation and superdense coding.

\subsection{Teleportation}

   As we discussed above, teleportation of a $one$-qubit state may be possible
with a W-state with even number of particles. For even number, the smallest 
number is two. A two-qubit W-state is really one of the Bell-states, 
$\ket{\psi^+} = \frac{1}{\sqrt{2}}(\ket{10} + \ket{01})$. So it can
be used for the teleportation of an unknown $one$-qubit state. First non-trivial
case is that of four-qubit W-state. We shall discuss the extension of the 
teleportation protocol for this case, then generalize the arguments for the
$n$-qubit state with even $n$.

The four-qubit W-state can be written as
\begin{eqnarray}
\ket{W_4}=\frac{1}{\sqrt{2}}(\ket{\psi^{+}} \ket{00} + \ket{00} \ket{\psi^{+}}).
\end{eqnarray}
We note that we can rewrite the four-qubit W-state in this way for any distribution
of particles. The entropy of any two-qubit subsystem is one. Clearly one can teleport
a two-qubit state $ \alpha \ket{00} + \beta \ket{\psi^+}$ \cite{pap}. This state
can be used to encode a $one$-qubit state. However our
interest is in the teleportation of a {\em genuine} $one$-qubit state. Let us consider this
situation and examine the possibility of teleporting the state $\ket{\psi}_a = \alpha \ket{0}
+ \beta \ket{1}$.
Suppose Alice has qubits `1' and `2' of the resource state and unknown qubit 'a' 
and Bob has qubits `3' and `4'. The combined input state is
\begin{eqnarray}
\ket{\psi}_{a}\ket{W_4}_{1234}=\frac{1}{\sqrt{2}}(
\alpha \ket{0}_{a}\ket{\psi^{+}}_{12}\ket{00}_{34} + 
\alpha \ket{000}_{a12}\ket{\psi^{+}}_{34} +
\beta \ket{1}_{a}\ket{\psi^{+}}_{12} \ket{00}_{34} +
\beta \ket{100}_{a12}\ket{\psi^{+}}_{34}).
\end{eqnarray} 
The above equation can be rewritten as
\begin{eqnarray}
\ket{\psi}_{a}\ket{W}_{1234} & = & \frac{1}{2}[\ket{\xi^{+}}_{a12}
(\alpha \ket{00}_{34}+\beta \ket{\psi^+}_{34})+
\ket{\xi^{-}}_{a12}(\alpha \ket{00}_{34}-\beta \ket{\psi^+}_{34})+ \nonumber \\
&  & \;\;\;\;\;\; \ket{\eta^{+}}_{a12}(\alpha \ket{\psi^+}_{34}+\beta \ket{00}_{34}) +
\ket{\eta^{-}}_{a12}(\alpha \ket{\psi^+}_{34}-\beta \ket{00}_{34})],
\end{eqnarray}
where we have used a set of  three-qubit orthogonal states
\begin{eqnarray}
 \ket{\xi^{\pm}} &
= & \frac{1}{\sqrt{2}}(\ket{0}\ket{\psi^{+}} \pm \ket{100}), \nonumber \\
\ket{\eta^{\pm}} &
= & \frac{1}{\sqrt{2}}(\ket{000}\pm \ket{1}\ket{\psi^{+}}).
\end{eqnarray}
Using these states, Alice makes three-particle projective measurements.
After these measurements, Bob's two qubits are in the states
$\alpha \ket{00}_{34} \pm \beta \ket{\psi^{+}}_{34}, 
\alpha \ket{\psi^{+}}_{34} \pm \beta \ket{00}_{34}.$

As we have discussed in the last section briefly,
now Bob can adopt at least three different strategies. In the first strategy, Bob
can apply suitable multinary unitary transformation on his qubits to convert the four
of the above states into $(\alpha \ket{00}_{34} + \beta \ket{\psi^+}_{34})$,
and take the view that the unknown $one$-qubit state is now encoded in
the subspace of the Hilbert space of his two qubits. In the second
strategy, Bob can apply a different multinary unitary transformations
and transfer the initial unknown qubit state to one of his qubits. In the
third strategy, he can bring a Bell-pair and apply a three-qubit measurement
to transfer the initial unknown qubit state to one of the qubits of the Bell-pair.
Let us now see how to implement these strategies. {\em First strategy:} This strategy
can be implements by the following set of multinary unitary transformations
\begin{eqnarray}
 U^{s1}_1 & = & \ket{00}\bra{00} + \ket{11}\bra{11} + \ket{\psi^+}\bra{\psi^+} + \ket{\psi^-}\bra{\psi^-},  \nonumber \\
 U^{s1}_2 & = & \ket{00}\bra{\psi^+} + \ket{11}\bra{11} + \ket{\psi^+}\bra{00} + \ket{\psi^-}\bra{\psi^-},  \nonumber \\
 U^{s1}_3 & = & \ket{00}\bra{\psi^+} + \ket{11}\bra{11} - \ket{\psi^+}\bra{00} + \ket{\psi^-}\bra{\psi^-},  \nonumber \\
 U^{s1}_4 & = & \ket{00}\bra{00} + \ket{11}\bra{11} - \ket{\psi^+}\bra{\psi^+} + \ket{\psi^-}\bra{\psi^-}.
\end{eqnarray}
The operator $U^{s1}_1$ is identity operator. We have used the basis set
$\{\ket{00}, \ket{11}, \ket{\psi^+}, \ket{\psi^-}\}$, where $\ket{\psi^\pm} =
\frac{1}{\sqrt{2}}(\ket{01} \pm \ket{10})$. These transformations, basically
represent the application of the transformations $\{ \sigma_{0}, \sigma_{1},
i \sigma_{2}, \sigma_{3} \}$ in the subspace $\{\ket{00}, \ket{\psi^+} \}$.
After Alice conveys her results to Bob using two classical bits, Bob could apply one  
of the above suitable transformations to convert the state of his two qubits 
to $(\alpha \ket{00}_{34} + \beta \ket{\psi^+}_{34})$, and take the view 
that the unknown state now resides in the Hilbert space of his two qubits.
{\em Second strategy:} In this strategy, Bob applies such multinary transformations
on his two qubits so that they get unentangled and the unknown state is transferred
to one of the his qubits. Such a set of multinary unitary transformations would be
\begin{eqnarray}
 U^{s2}_1 & = & \ket{00}\bra{00} + \ket{11}\bra{11} + \ket{01}\bra{\psi^+} + \ket{10}\bra{\psi^-},  \nonumber \\
 U^{s2}_2 & = & \ket{00}\bra{\psi^+} + \ket{11}\bra{11} + \ket{01}\bra{00} + \ket{10}\bra{\psi^-},  \nonumber \\
 U^{s2}_3 & = & \ket{00}\bra{\psi^+} + \ket{11}\bra{11} - \ket{01}\bra{00} + \ket{10}\bra{\psi^-},  \nonumber \\
 U^{s2}_4 & = & \ket{00}\bra{00} + \ket{11}\bra{11} - \ket{01}\bra{\psi^+} + \ket{10}\bra{\psi^-} .
\end{eqnarray}
The structure of these transformations is determined by the fact that we would like
the following transformation by unitary operators -- $\{\ket{00}, \ket{\psi^+} \}
\to \{\ket{00}, \ket{01} \}$.
These transformations would convert the Bob's two qubits to the unentangled
state $\ket{0}(\alpha \ket{0} + \beta \ket{1}).$ In this way,  the input state
has been transmitted to one of the Bob's qubits. {\em Third strategy:} In this 
strategy of {\em serial} teleportation, Bob uses another Bell-pair and 
another three-qubit measurement to
transfer the input state to one of the qubits. This strategy works as follow. 
Bob brings a Bell-pair $\frac{1}{\sqrt{2}}(\ket{00}+\ket{11})_{56}$ and performs a joint measurement on two qubits in his possession and one qubit of the Bell-pair with suitable measurement states. After this he carrys out unitary operation on the other qubit 
to convert it into the state of the unknown qubit. Let us illustrate this procedure with an
example. Let the Bob's qubit be in the $ \alpha \ket{00}_{34}+\beta \ket{\psi^+}_{34}$ state. 
We can rewrite this state with that of the Bell-pair as
\begin{equation}
\frac{1}{\sqrt{2}}(\ket{000}_{345}\alpha \ket{0}_{6}+\ket{001}_{345}\alpha \ket{1}_{6}+
\ket{\psi^{+}}_{34}\ket{0}_{5}\beta \ket{0}_{6}+\ket{\psi^{+}}_{34}\ket{1}_{5}\beta \ket{1}_{6}).
\end{equation}
If Bob perform measurement using the orthogonal states
\begin{eqnarray}
\ket{\phi_{1}^{\pm}} & = & \frac{1}{\sqrt{2}}(\ket{000}\pm \ket{\psi^{+}} \ket{1}), \nonumber \\ 
\ket{\phi_{2}^{\pm}} & = & \frac{1}{\sqrt{2}}(\ket{001}\pm \ket{\psi^{+}} \ket{0}),
\end{eqnarray}
then we can write above equation $(33)$ as
\begin{equation}
\frac{1}{2}(\ket{\phi_{1}^{+}}_{345}(\alpha \ket{0}+\beta \ket{1})_{6}
+(\ket{\phi_{1}^{-}}_{345}(\alpha \ket{0}-\beta \ket{1})_{6} \nonumber \\
+(\ket{\phi_{2}^{+}}_{345}(\beta \ket{0}+\alpha\ket{1})_{6}
+(\ket{\phi_{2}^{+}}_{345}(\beta \ket{0}-\alpha\ket{1}))_{6}.
\end{equation}
Then Bob applies suitable Pauli operators $\{\sigma_{0}, \sigma_{1}, i \sigma_{2},
\sigma_{3}\}$ on the qubit '6' to obtain the original qubit. Similar procedure will
work on other three of Bob's two-qubit states.

One can immediately generalize the above strategies to an $n$-qubit state, where $n$ is even.
This is easy to see because any such state $\ket{W_{2n}}$ of $2n$ qubits can be written as
\begin{equation}
\ket{W_{2n}}=\frac{1}{\sqrt{2}}(\ket{000...0}_{n}\ket{W_n}+\ket{W_n}\ket{000...0}_{n}).
\end{equation}.
This state has structure similar to the $\ket{W_4}$ state. It can be used to
teleport a state of the type $\alpha \ket{000...0} + \beta \ket{W_n}$. To teleport
a genuine $one$-qubit state using this resource, one can use all the three strategies
discussed above with suitable mappings.  

\subsection{Superdense coding }

    As we have seen that for even number of qubits, one specific partition has entropy
 one. This happens when Bob and Alice, each has half of the qubits. It has been
 shown \cite{pap} that for four-qubit W-state, with this kind of sharing Alice
 can transmit three classical bits to Bob by sending two qubits. One set of
 multi-unary transformations on the $\ket{W_{4}}$ which are useful is:
 $\{\sigma_{0} \otimes \sigma_{0}, \sigma_{0} \otimes \sigma_{1}, \sigma_{1} \otimes i \sigma_{2}, \sigma_{1} \otimes \sigma_{3}, i \sigma_{2} \otimes \sigma_{0}, i \sigma_{2} \otimes \sigma_{1}, \sigma_{3} \otimes i \sigma_{2}, \sigma_{3} \otimes \sigma_{3}  \}$. 
Maximal superdense
 coding \cite{pb} is not possible with such a state. In a more general scenario of
 $2n$-qubits, Alice can apply multi-unary transformations to her $n$ qubits and
 send the $n$ qubits to Bob. Bob can then recover $n + 1$ classical bits. 

\section{Conclusion}

   We have considered generalized W-states, $\ket{WM_n}$.
    We have obtained the conditions for the successful teleportation and
 superdense coding with such a state.  For
 a given number of qubits, there exist many such W-states which
can help us with the tasks. With these states, one can teleport
a $n \over 2$-qubit (or $(n-1) \over 2$-qubit state when $n$ is odd) state which
has {\em only}  two suitable terms. One cannot teleport a state with larger
number of terms.  For superdense coding, one can gain at most
one classical bit, i.e., with suitable states, one can transmit
{\em at most} $(m+1)$ classical bits by sending $m$ qubits.
Therefore maximal teleportation and superdense coding are not
possible with this state. 
The condition for the partition of the qubits that help
in the teleportation or superdense coding implies that the von Neumann 
entropy is one for the partitioned subsystems. We have also shown using
 any of these generalized W-states, one would be able to teleport the unknown
 state of a {\em one}-qubit by any of the three methods. In the first
 method the state would appear in the superposition of the two states
 of a higher dimension Hilbert space system. In the second method,
 we can apply multinary unitary transformation to transfer the state to
 one of the qubits of Bob. In the third method, Bob can use another
 Bell pair and suitable measurements to transfer the unknown state
 to one of the qubits of the Bell pair. We have shown the implementation
of these strategies explicitly for the case of four-qubit W-state, which
is a special case of generalized W-state in the case of four-qubits.

\end{document}